\documentclass{PoS}

\newcommand{\sigmaSI}{\sigma_{\rm SI}}

\newcommand{\ifb}{\rm fb^{-1}}
\newcommand{\ipb}{\rm pb^{-1}}

\newcommand{\gev}{{\rm GeV}}
\newcommand{\tev}{\rm TeV}
\newcommand{\pb}{\rm pb}

\newcommand{\eqref}[1]{Eq.~(\ref{#1})}

\newcommand{\lsim}{\lower.7ex\hbox{$\;\stackrel{\textstyle<}{\sim}\;$}}
\newcommand{\gsim}{\lower.7ex\hbox{$\;\stackrel{\textstyle>}{\sim}\;$}}

\newcommand{\be}{\begin{equation}}
\newcommand{\ee}{\end{equation}}
\newcommand{\Dsle}[1]{\slash\hskip -0.28 cm #1}
\newcommand{\met}{{\Dsle E_T}}
\newcommand{\Dslp}[1]{\slash\hskip -0.23 cm #1}
\newcommand{\mpt}{{\Dslp p_T}}
\newcommand{\bea}{\begin{eqnarray}}
\newcommand{\eea}{\end{eqnarray}}

\title{New Signatures of WIMPless Dark Matter }

\ShortTitle{New Signatures of WIMPless Dark Matter}

\author{\speaker{Jason Kumar}\\
        University of Hawaii\\
        E-mail: \email{jkumar@hawaii.edu}}

\abstract{We consider new signatures of WIMPless dark matter, particularly
those which can be used to test models with low dark matter mass.  We focus on detection
prospects at hadron colliders through the production of new heavy
QCD-coupled particles,
which decay to WIMPless dark matter plus jets.  We find that the Tevatron can
probe a significant fraction of the low mass parameter space with data already
taken, and the LHC will have even better detection prospects with its first physics
run.  }

\FullConference{Identification of Dark Matter 2010-IDM2010\\
		July 26-30, 2010\\
		Montpellier France}

\begin{document}

\section{Introduction}

There has been interest in low mass models of dark matter, especially
since the DAMA collaboration reported 8.9$\sigma$ evidence of an annual
modulation in their event rate~\cite{Bernabei:2010mq}, which could be consistent with dark
matter with a low-mass ($m_X \sim {\cal O}(10)~\gev$) and large spin-independent scattering
cross-section ($\sigmaSI \sim 10^{-2-5}~\pb$).  This interest has been further
heightened by recent data from other experiments.
The CoGeNT experiment reported an unexplained event rate which is also consistent with
a low-mass, large $\sigmaSI$ dark matter model~\cite{Aalseth:2010vx,Hooper:2010uy}.
Finally, the CRESST collaboration has
reported at this meeting a preliminary excess of events in the oxygen band, which could
also be consistent with scattering of a low-mass dark matter particle~\cite{CRESST}.  A
great question surrounding all of these signals is how they can be distinguished from
an unknown background.

However, there have been contrary data from Xenon100~\cite{Aprile:2010um}
and from CDMS~\cite{CDMS}, which are in tension with these low-mass signals.  A new analysis
of bounds from XENON10 and XENON100~\cite{Savage:2010tg} places constraints on
low-mass dark matter as well.  A
new preliminary analysis of Xenon10 data, presented at this conference by Peter
Sorensen~\cite{PSorensen}, also exhibits significant tension with this data.  However, even
these analyses come with significant controversy~\cite{xenonquestions}.

The basic source of this controversy is the measurement of low recoil energies.  Low mass
dark matter tend to transfer less momentum to Standard Model nuclei during scattering, resulting in
less recoil energy.  For $m_X \lsim 10~\gev$, the recoil energies are near
the threshold for many dark matter detectors.  Essentially, the controversy surrounds whether or
not experiments such as CDMS or XENON10/100 are truly sensitive to such low recoil energies.

It seems fair to say that the experimental status of low-mass dark matter is in a state
of flux.  In the face of this conflicting data one might be led to wonder, ``What is a simple
theorist to do?"  In these proceedings, we will avoid the question of whether
the experiments are actually seeing the interactions of a low-mass dark matter
particle.  Instead we will focus
on two other questions:
\begin{itemize}
\item{What theoretical models could yield a dark matter particle with low mass
and large spin-independent scattering cross-section? }
\item{What experiments could check this result, preferably with data which
has either already been taken or will be taken soon?}
\end{itemize}
To answer the first question, we will focus on WIMPless dark matter models~\cite{Feng:2008ya}.
Since the questions surrounding the experimental evidence arise from the problems
which direct detection experiments often
have in measuring low recoil energies, it is advisable to check these signals
with methods very different from direct detection,
and which do not face the difficulty of low recoil energy thresholds.

It was pointed out in \cite{Hooper:2008cf,Feng:2008qn} that Super-Kamiokande is a very good experiment
for checking these low-mass signals through searches for dark matter annihilating in the
sun.  Firstly, the neutrino flux due to dark matter annihilation is determined by solar
dark matter capture rate, which is subject to fewer astrophysics uncertainties than other
indirect detection search strategies.  Secondly, Super-Kamiokande is sensitive to very
low-energy neutrinos, thus allowing it to probe the annihilation of low-mass dark matter.
In particular, it was shown in \cite{Feng:2008qn} that Super-Kamiokande could probe this
low-mass regime with the sample of fully-contained muons already taken in SK run III.

In these proceedings, I will discuss another search strategy which is poised to probe low
mass dark matter models with data already taken or soon to be available: collider experiments.

\section{WIMPless dark matter}

WIMPless dark matter is a hidden sector dark matter candidate which automatically
has approximately the correct thermal relic density to explain our cosmological
observations of dark matter, generalizing the ``WIMP Miracle" beyond the case of
weak interactions at the electroweak symmetry breaking scale \cite{Feng:2008ya}.
In this model, both the
hidden sector and MSSM sector receive SUSY-breaking through gauge-mediation from
a single SUSY-breaking sector.  The physics of the SUSY-breaking sector thus sets
the soft breaking mass scale of the hidden sector and of the MSSM sector (which in
turn sets the weak scale) through the relation
\be
{g_{X}^2 \over m_X }, {g_{weak}^2 \over m_{weak} } \propto {M_{mess.} \over F}
\ee
where $g_X$ and $m_X$ are the coupling and soft SUSY-breaking scale of the hidden
sector, $F$ is the SUSY-breaking $F$-term vev, and $M_{mess.}$ is the mass scale
of the messengers.  This relation is important because the thermal relic density
is set by the annihilation cross-section, $\rho \propto \langle \sigma_{ann.} v
\rangle^{-1}$, which in turn is determined by the
relation  $\langle \sigma_{ann.} v \rangle \propto {g^4 \over m^2}$.
A stable particle at the hidden sector soft scale (the WIMPless
candidate) then necessarily
has approximately the same relic density as a WIMP.

The interesting feature of the model is that the relic density is very robust, and
is independent of the details of the hidden sector.  In particular, the model has
approximately the correct density for a wide possible range of dark matter candidate
mass, and regardless of whether the particle is a boson or fermion.
This opens up the window for experimental search strategies beyond those typically used
for WIMPs.

If the WIMPless particle is a scalar, then it can have a much larger $\sigmaSI$ than
one would expect from neutralino WIMPs.  The basic reason is that scalar
dark matter can couple to the Standard Model through a dimension 5 operator,
$(X^* X)(\bar f f)$, while a Majorana fermion must couple through a
dimension 6 operator, such as $(\bar \chi \chi)(\bar f f)$.  This suggests that
a scalar WIMPless candidate is a good candidate for matching these low
mass, large $\sigmaSI$  experimental signals.

WIMPless dark matter can couple to the SM sector through Yukawa couplings involving
a 4th generation quark.  In~\cite{Feng:2008dz}, a specific model was proposed to
match the DAMA signal.  In this model, dark matter couples exclusively to third generation
quarks\footnote{This is a simple way to satisfy
flavor-changing neutral current constraints, while satisfying the observed Yukawa hierarchy,
in which each generation mixes largely with the next generation.} via the Yukawa coupling
\be
V = \lambda_L X \bar Q'_{L(T',B')} q_{L(t,b)} + \lambda_{Rt} X \bar T'_R t_R +
\lambda_{Rb} X \bar B'_R b_R + h.c.
\ee
Gauge-invariance implies that the $Q'_{L} = \{T'_L,B'_L \}$, $T'_R$, $B'_R$ form
a mirror 4th generation multiplet.  This multiplet must also be charged under the
same hidden sector symmetry which stabilizes $X$, so it is really an exotic 4th generation
quark multiplet.  The mass of the 4th generation
quarks is constrained by a combination of direct searches, perturbativity and
precision electroweak constraints~\cite{Kribs:2007nz} to the the
range $300~\gev \lsim m_{T',B'} \lsim 600~\gev$.

This WIMPless model couples to Standard Model nuclei through a loop of b quarks,
which couples to the gluon content of the nucleus.
To match these low mass signals with $300~\gev \leq m_{T',B'} \leq 600~\gev$ and
$m_X = 1-10~\gev$, one would need $\lambda_{(L,R)b} \sim 0.3-1$~\cite{Feng:2008dz}.

\section{Tests at hadron colliders}

Hadron colliders excel at the production of particles which couple
to QCD.  For this reason, search strategies at hadron colliders
often focus on the production of
strongly coupled particles (such as the gluino), which then decay to
other particles (such as neutralinos).  This strategy is especially
important for dark matter, which is necessarily decoupled from QCD.
Instead, hadron colliders produce heavy QCD-coupled particles which
are also charged under a new symmetry, and then decay to the lightest
particle charged under the same new symmetry, which is dark matter.

The simplest such decay we can write is $\phi_{heavy} \rightarrow
\phi_{SM}X$, where $\phi_{heavy}$ is the the new heavy particle produced
directly at the collider, $\phi_{SM}$ is a Standard Model particle which
carries the same QCD quantum numbers as $\phi_{heavy}$, and $X$ is the
QCD-neutral dark matter particle.
Since the dark matter is
necessarily long-lived and does not couple to strong or electromagnetic
forces, this results in a collider signature of missing
transverse energy.
Collider experiments are thus ideally suited for testing low-mass
dark matter models, as the event rate increases for smaller
dark matter mass.

Supersymmetry analyses focus on the case where $\phi_{heavy}$ is
either a color-octet fermion or a color-triplet scalar, and $m_X \sim
100-1000~\gev$.
WIMPless dark matter provides another simple example of this general set-up, where
$\phi_{heavy} =T',B'$ is a color-triplet fermion, which decays by
$\phi_{heavy} \rightarrow b,t +X$.  A general analysis of
this possibility for both the Tevatron and LHC was performed in~\cite{Alwall:2010jc}.

The process studied was the production of the up-type exotic quark
through the process $pp(\bar p) \rightarrow T' \bar T' \rightarrow XXt \bar t$.
The signature searched for is missing energy, plus a large number of jets arising from
top decay.  Although the semi-leptonic decay channel was also studied, we found that the
most promising channel was fully hadronic, since a lepton veto also rejects most Standard
Model backgrounds with missing energy from neutrinos.

Initial cuts for the hadronic channel are
\begin{itemize}
\setlength{\itemsep}{1pt}\setlength{\parskip}{0pt}\setlength{\parsep}{0pt}
\item No isolated electrons, muons or tau-tagged jets with $|p_T^l| >
2~\gev$.
\item Minimum missing transverse energy: $\met > 100~\gev$.
\item At least 5 jets with $|p_T^j| > 20~\gev$ (Tevatron) or $|p_T^j| >
40~\gev$ (LHC).
\item Minimum $\Delta\phi(\mpt, p_T^j)$ for the leading jets:
$\Delta\phi(\mpt, p_T^{j1}) > 90^\circ$ and $\Delta\phi(\mpt,p_T^{j2}) >
50^\circ$ (Tevatron); $\Delta\phi(\mpt, p_T^{j}) > 11.5^\circ$ for
the first, second and third leading jets (LHC).
\end{itemize}
Additional cuts on $~\met$, the number of jets, and $H_T$ (defined by
$H_T = \sum_{i=1}^5 |p_T^j|$) are imposed point by point to maximize signal
significance.

The main Standard Model backgrounds are $t\bar t$, leptonically-decaying
$W\,{\rm + jets}$ with a missed lepton or where a $\tau$ has been mistagged
as a jet, and $Z\,{\rm + jets}$ where $Z \rightarrow \bar \nu \nu$.
The $\Delta \phi (\mpt, p_T )$ cuts effectively eliminate the QCD multi-jet background.
The signal and
these backgrounds were simulated with Madgraph/MadEvent - Pythia 6.4.20 - PGS4~\cite{simulation}.

\begin{figure}[tb]
\begin{center}
\includegraphics*[width=0.48\columnwidth]{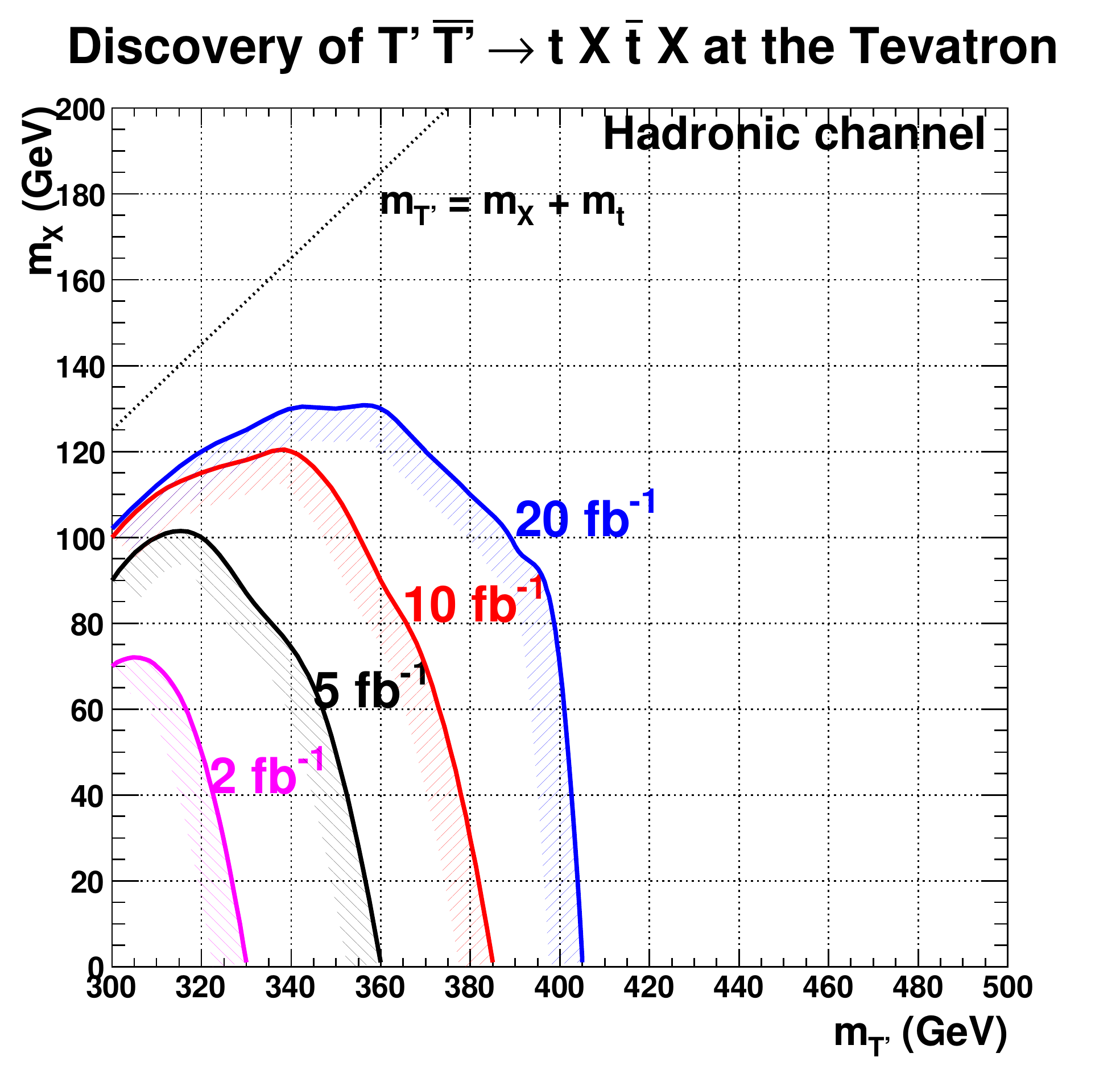}
\includegraphics*[width=0.48\columnwidth]{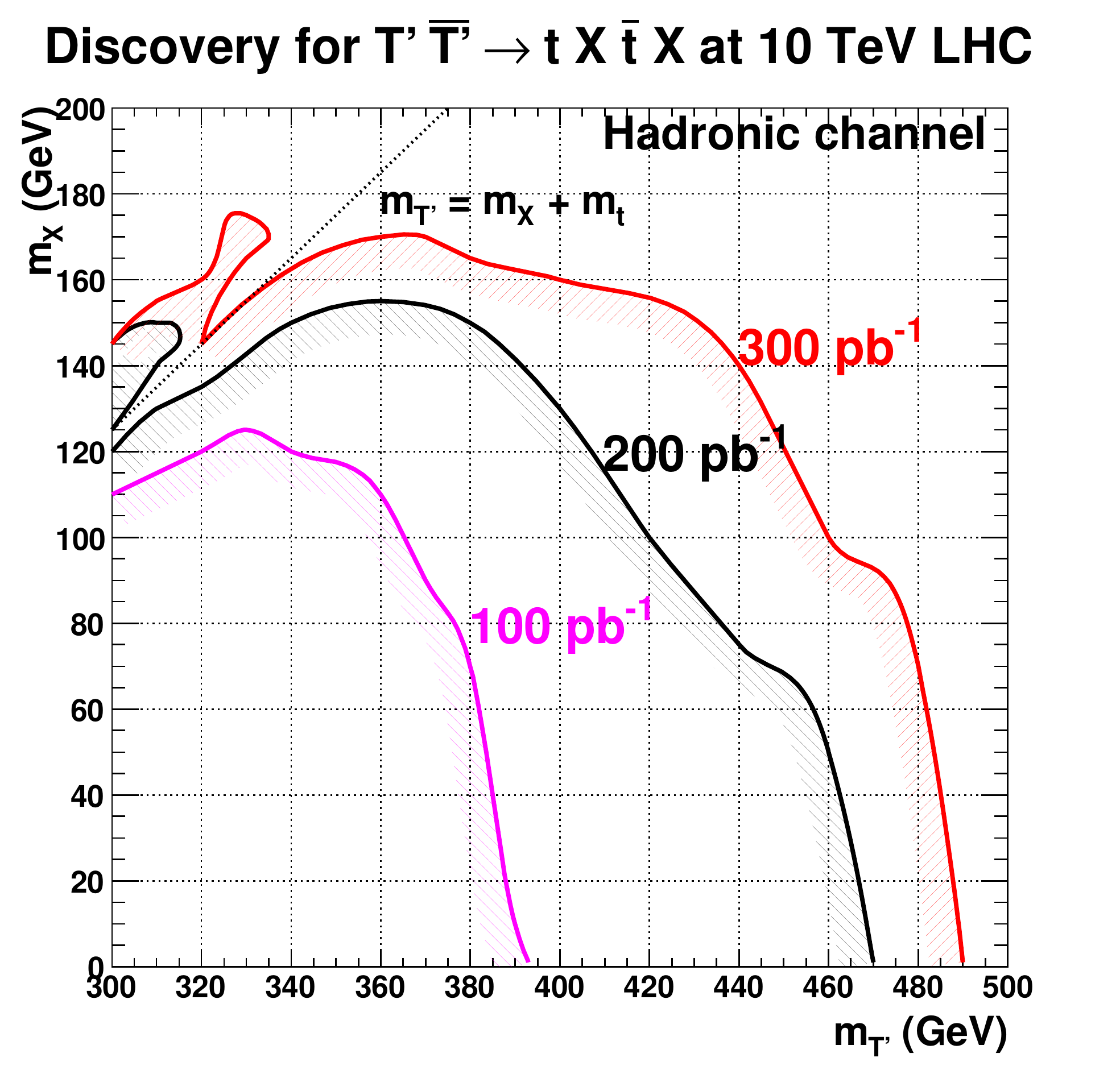}
\end{center}
\vspace*{-.25in}
\caption{\label{fig:detection} 3$\sigma$ (Gaussian equivalent) evidence contours
for the hadronic channel for the Tevatron (left) with integrated luminosities
2, 5, 10, and $20~\ifb$ and for the LHC (right) with integrated luminosities
100, 200, and $300~\ipb$. For each point in
parameter space, the cut with the best significance has been chosen.}
\end{figure}

Detection prospects for the Tevatron and for the LHC at $10~\tev$ center-of-mass
energy are plotted in Figure~\ref{fig:detection}.
The detection prospects for $300~\ipb$ at the LHC at $10~\tev$ roughly correspond to
$1~\ifb$ for a $7~\tev$ run.  We thus see that the LHC can probe a significant fraction
of the allowed parameter space in its first physics run.

\section{Conclusion}  We have found that hadron colliders can provide interesting
new constraints on low mass models of dark matter which could match the data of
DAMA, CoGeNT and CRESST, including WIMPless dark matter (see also~\cite{hadrontests}).
The data and constraints
in this region of parameter-space are constantly changing; indeed, during this conference
we have seen a new preliminary analysis which can place tight constraints on
this region of parameter space.  As pointed out in~\cite{Chang:2010yk}, models
which couple differently to protons and neutrons can potentially explain the low mass
data while remaining consistent with even these new constraints.  Since WIMPless models
couple to the Standard Model via Yukawa couplings, it is easy to construct models
whose couplings violate isospin (indeed, isospin violation should perhaps be expected).
A study of such models is in preparation~\cite{isospinWIMPless}, and hadron colliders would
again be an ideal venue for testing such models.

\acknowledgments

We are grateful to the organizers of IDM2010, and to J.~Alwall, J.~L.~Feng, and
S.~Su for discussions and collaboration.  JK is supported by DOE grant DE-FG02-04ER41291.

\end{document}